\documentclass[aps,prb,twocolumn,A4paper,showpacs,showkeys]{revtex4}

\usepackage{graphicx}
\usepackage{amssymb}

\DeclareGraphicsExtensions{.eps}
\begin{document}

\title{Microcavity exciton-polariton mediated Raman scattering: experiment
and theory}
\author{A. Bruchhausen}
\email{bruchhau@cab.cnea.gov.ar}
\affiliation{Instituto Balseiro \& Centro At\'omico Bariloche, C.N.E.A., R8402AGP
Bariloche, RN, Argentina.}
\author{L. M. Le\'on Hilario}
\affiliation{Instituto Balseiro \& Centro At\'omico Bariloche, C.N.E.A., R8402AGP
Bariloche, RN, Argentina.}
\author{A. A. Aligia}
\altaffiliation[Also at ]{CONICET, Argentina.}
\affiliation{Instituto Balseiro \& Centro At\'omico Bariloche, C.N.E.A., R8402AGP
Bariloche, RN, Argentina.}
\author{A. M. Lobos}
\altaffiliation[Present address: ]{DPMC-MaNEP, Universit\'e de Gen\`eve, 24 Quai Ernest Ansermet, 1211 Gen\`eve
4, Switzerland.}
\affiliation{Instituto Balseiro \& Centro At\'omico Bariloche, C.N.E.A., R8402AGP
Bariloche, RN, Argentina.}
\author{A. Fainstein}
\altaffiliation[Also at ]{CONICET, Argentina.}
\affiliation{Instituto Balseiro \& Centro At\'omico Bariloche, C.N.E.A., R8402AGP
Bariloche, RN, Argentina.}
\author{B. Jusserand}
\affiliation{Institut des NanoSciences de Paris, UMR CNRS 7588, Universit\'e Pierre et
Marie Curie, Campus Boucicaut, 140 Rue de Lourmel, 75015 Paris, France.}
\author{R. Andr\'e}
\affiliation{Laboratoire de Spectrom\'etrie Physique, Universit\'e Joseph
Fourier-Grenoble 1, CNRS, Bo\^{\i}te Postale 87, F-38402, St. Martin d'Heres
Cedex, France.}

\begin{abstract}
We studied the intensity of resonant Raman scattering due to optical phonons in a planar II-VI type semiconductor microcavity in the regime of strong coupling between light and matter. Two different sets of independent experiments were performed at near outgoing resonance with the middle polariton (MP) branch of the cavity. In the first, the Stokes-shifted photons were kept at exact resonance with the MP, varying the photonic or excitonic character of the polariton. In the second, only the incoming light wavelength was varied, and the resonant profile of the inelastic scattered intensity was studied when the system was tuned out of the resonant condition. Taking some matrix elements as free parameters, both independent experiments are quantitatively described by a model which incorporates lifetime effects in both excitons and photons, and the coupling of the cavity photons to the electron-hole continuum. The model is solved using a Green's function approach which treats the exciton-photon coupling nonperturbatively.
\end{abstract}

\pacs{78.30.Fs, 78.67.-n, 71.36.+c}
% 78.30.Fs Infrared and Raman spectra (III-V and II-VI semiconductors)
% 78.67.-n Optical properties of low-dimensional, mesoscopic, and nanoscale materials and structures
% 71.36.+c Polaritons (including photon-phonon and photon-magnon interactions)
%% 63.20.kk Phonon interactions with other quasiparticles

\keywords{Raman spectroscopy, II-VI semiconductors, optical microcavities,
optical confinement, cavity exciton-polaritons}
\date{\today}
\maketitle

%%%%%%%%%%%%%%%%%%%%%%%%%%%%%%%%%%%%%%%%%%%%%%%%%%%%%%%%%%%%%%%%%%%%%%%%%%%%%%%%
%%%%%%%%%%%%%%%%%%%%%%%%%%%%%%%%%%%%%%%%%%%%%%%%%%%%%%%%%%%%%%%%%%%%%%%%%%%%%%%%
\section{Introduction and Motivation}
%%%%%%%%%%%%%%%%%%%%%%%%%%%%%%%%%%%%%%%%%%%%%%%%%%%%%%%%%%%%%%%%%%%%%%%%%%%%%%%%
%%%%%%%%%%%%%%%%%%%%%%%%%%%%%%%%%%%%%%%%%%%%%%%%%%%%%%%%%%%%%%%%%%%%%%%%%%%%%%%%

Semiconductor microcavities (MC's) are micron-scaled photonic structures, consisting of two planar distributed Bragg reflectors (DBR's) separated by a spacer with a thickness of the order of the photon wavelength, forming a high quality Fabry-P\'erot resonator, and leading to a localized photonic mode inside the MC.\cite{Book-ConfElAndPhot(95)} This spacer usually embeds quantum wells (QW's) conveniently located at the antinodes of the optical-cavities electric field. The strong vertical confinement of both the electric field and the QW excitons result in a strong and by design controllable light-matter interaction. When this MC mode is resonant with the excitonic transitions, and when the life time broadening of the MC-photons ($\gamma _{f}$) and QW-excitons ($\gamma _{e}$) are smaller than the photon-exciton interaction ($\Omega$), the system is under the strong-coupling regime, where coherent Rabi oscillations between the bare states occur. This regime is evidenced by the characteristic anticrossing behavior exhibited by the exciton and the cavity mode when they are brought into resonance. As a consequence, the ``new'' eigenstates of the system are strongly mixed exciton-photon states, \emph{cavity-polaritons}.\cite{Book-ConfElAndPhot(95), Libro-CavityPolaritons(03), SemicondSciTech18-issue10(03), JPhysCondensMatter18-SpecialIssue} These quasi-particles have shown to exhibit a large number of interesting properties, most of them arising from their inherited bosonic character. This character together with their very small effective mass (typically 10$^{-4}$ times smaller than that of the bare excitons) enables processes like the condensation into a quantum degenerate Bose-Einstein state\cite{Kasprzak-Nature443-409(06), Balili-Science316-1007(07)} in thermal equilibrium with the lattice,\cite{Deng-PRL97-146402(06)} and the \emph{boser} effect or stimulated emission,\cite{Imamoglu-PLA214-193(96), Senellart-Bloch-PRL82-1233(99), Dang-PRL81-3920(98)} among other non-linear phenomena.\cite{Book-PhysicsofSemiconductorMicrocavities(06)} This made the cavity-polariton systems not only very interesting from the basic fundamental point of view, but also very promising in the field of technological application in coherent light emitting devices.\cite{Libro-CavityPolaritons(03), SemicondSciTech18-issue10(03), JPhysCondensMatter18-SpecialIssue, Book-PhysicsofSemiconductorMicrocavities(06), deLima-PRL7545326(06)} Efforts have been devoted incessantly to the study of cavity-polaritons experimentally and theoretically during the last fifteen years,\cite{Libro-CavityPolaritons(03), SemicondSciTech18-issue10(03), JPhysCondensMatter18-SpecialIssue} since the first experimental observation.\cite{Weisbuch-PRL67-3314(92)} \\

The process of inelastic light scattering due to vibrations in semiconductors and in microcavities for the case where the energy of the light mode is far from the corresponding one of electronic excitations, is usually very well described by treating the light-matter coupling perturbatively.\cite{Cardona-LightScattering-II, Fainstein-PRL86-3411(01)} In this case the Raman effect is mediated by electronic or excitonic states by a second order process in exciton-photon interaction. 
Instead, in microcavities when the cavity mode is tuned near the QW-exciton energy the perturbative approach fails.\cite{Fainstein-PRL78-1576(97), Tribe-PRB56-12429(97), Fainstein-PRB57-R9439(98), Bruchhausen-PRB68-205326(03), Stevenson-PRB67-81301(R)(03)} The exact treatment of the light-matter interaction, i.e. the cavity exciton-polariton picture, is essential and the inelastic process (now mediated by polaritons) must be described by a first order perturbation theory of the polariton-phonon interaction.\cite{Fainstein-PRL78-1576(97), Fainstein-PRB57-R9439(98), Weisbuch-LightScattering-III}

Since the Raman process due to optical phonons is a coherent process, it is very sensitive to the damping of the resonant state.\cite{Fainstein-PRL78-1576(97), Tribe-PRB56-12429(97), Fainstein-PRB57-R9439(98), Bruchhausen-PRB68-205326(03)} Resonant Raman scattering (RRS) experiments as a function of photon-exciton detuning evidence the dependence of the polariton--phonon and polariton--external-photon coupling with the mode mixing in the process. This is distinctly demonstrated by the maximum in the scattering efficiency near the mode anticrossing (zero detuning) and the decreasing efficiency when the polariton character approaches either the pure exciton or pure photon state. \cite{Fainstein-PRB57-R9439(98), Bruchhausen-PRB68-205326(03)} Although existing simplified theoretical models describe the phenomenological and qualitative behavior of the experimental results, the relevance of rigorously including polariton dephasing and dumping effect has been strongly emphasized.\cite{Tribe-PRB56-12429(97), Bruchhausen-PRB68-205326(03), Bruchhausen-AIPCP772-1117(05)}

In this paper we present detailed RRS experimental results together with a theoretical improvement of the existing models that rigorously takes polariton lifetimes into account. The experiments are performed on II-VI semiconductor based microcavities and consist essentially of two types: the first showing the behavior of the Raman intensity as a function of cavity-photon--exciton detuning at exact resonance, and the second showing the behavior of the outgoing scattered process for a fixed detuning, and performing a scan with the excitation energy. Experiments are compared to the developed theory that uses a Green's function approach to include the damping effects and the electron-hole (free-exciton) states in a formal way. Using the matrix elements in the theory as free parameters, the theory describes accurately the Raman intensity simultaneously for both types of experiments. Bare life time broadening are obtained from a fitting procedure and the effects of the exciton lifetime in relation to the cavity-polariton lifetime are discussed.
%A description of the theory aimed to describe only the first type of experiments described above has been presented previously, \cite{LeonHilario-JPCondMatter19-176210(07),LeonHilario-InPress-(07)} but the Raman scans obtained by the second type of experiments provide additional severe constraints to the unknown parameters of the theory.  
An important improvement to the previously presented theoretical approaches\cite{LeonHilario-JPCondMatter19-176210(07), LeonHilario-InPress-(07)} to describe the experimental observations quantitatively is obtained in the present work by accounting also for the second type of experiment, which provides additional severe constrains to the unknown parameters of the theory.\\

The paper is organized as follows: In section~\ref{Experimental Details} the samples are shortly described and the experiments are presented. In section~\ref{Experimental Results} the Experimental results are shown. The theoretical model is introduced in section~\ref{Theory}: a short description of the existing phenomenological approaches is given, and afterwards the model Hamiltonian  and Green's function approach for the Raman intensity are explained. In section~\ref{Analisys and discussion} the experimental and calculated results are analyzed and discussed. Finally the conclusions of this work are summarized.

%%%%%%%%%%%%%%%%%%%%%%%%%%%%%%%%%%%%%%%%%%%%%%%%%%%%%%%%%%%%%%%%%%%%%%%%%%%%%%%%
%%%%%%%%%%%%%%%%%%%%%%%%%%%%%%%%%%%%%%%%%%%%%%%%%%%%%%%%%%%%%%%%%%%%%%%%%%%%%%%%
\section{Experimental Details}\label{Experimental Details} 
%%%%%%%%%%%%%%%%%%%%%%%%%%%%%%%%%%%%%%%%%%%%%%%%%%%%%%%%%%%%%%%%%%%%%%%%%%%%%%%%
%%%%%%%%%%%%%%%%%%%%%%%%%%%%%%%%%%%%%%%%%%%%%%%%%%%%%%%%%%%%%%%%%%%%%%%%%%%%%%%%

Experiments are all performed on a II-VI semiconductor microcavity. CdTe-based cavities are interesting since excitonic binding energies in CdTe quantum wells are much larger ($\sim 25\,$meV) as compared to GaAs ($\sim 10\,$meV). This is a consequence of a stronger electron-hole attraction that makes the excitonic states in this compounds much more stable. In addition, the ionic character of these compounds leads to a stronger light-matter coupling and hence to much larger Rabi gaps and stronger polaritonic effects compared to the GaAs-based structures. On the other hand Raman efficiency due to optical phonons is also larger allowing clearer studies, specially when the Stokes-shifted photon is tuned at near resonance to a polaritonic state, where luminescence hinders its observation.

The studied cavity was grown (100)-oriented by molecular beam epitaxy, and consists of two distributed Bragg reflectors (DBRs) enclosing a $\lambda /2$ Cd$_{0.4}$Mg$_{0.6}$Te spacer. Three centered CdTe quantum wells of nominal width $72\,$\AA\ and separated by $69\,$\AA\ from each other, are embedded in the spacer. The cavity mirrors consisted of 15.5 periods of Cd$_{0.4}$Mg$_{0.6}$Te / Cd$_{0.75}$Mg$_{0.25}$Te $\lambda /4$ layers for the top DBR, and 21 pairs for the DBR at the bottom, to balance the effect of the air-substrate asymmetry. 
The layers were grown in such a way that their thickness decreases in one direction (like in a wegde), and the cavity mode can thus be tuned by displacing the position of the laser spot on the sample. Experiments were performed at back-scattering $z(x,x')z'$ geometry, where $z$($z'$) and $x$($x'$) describe the incident(scattered) beam direction and light polarization, respectively.\cite{Cardona-FundamentalsOfSemicond(96)} Here, $x$ and $x'$ correspond to the (110) direction, $z$ to the (001) epitaxial growth direction, and since the configuration is back-scattering, $z'\equiv \bar{z}$. A tunable cw Ti:sapphire laser was used as excitation source at near normal incidence, with power $\lesssim 50$ $\mu$W, and focalized to a spot of $\varnothing \sim 50$ $\mu$m. The fixed collection cone normal to the sample surface was $\sim 3^{o}$, corresponding to an uncertainty in the in-plane wave vector of $k_{\parallel }\approx 4.2\times 10^{3}\,$cm$^{-1}$. A liquid-He cryostat was used to set the temperature at $4.5\,$K. At this temperature, the sample shows a double anticrossing of the cavity mode with both the $E1H1(1s)$ ($X_{1s}$) and the $E1H1(2s)$ ($X_{2s}$) heavy-hole exciton modes. 
%%%%%%%%%%%%%%%%%%%%%%%%%%%%%%%%%%%%%%%%%%%%%%%%%%%%%%%%%%%%%%%%%%%%%%%%%%%%%%%
\begin{figure}[t]
\begin{center}
\includegraphics*[keepaspectratio=true, clip=true, angle=0, width=0.7\columnwidth]{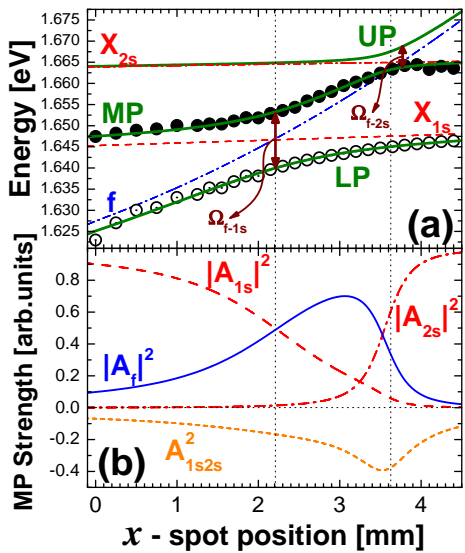}
\end{center}
\caption{(Color online) \textbf{(a)} Dispersions of the cavity-polariton modes as functions of the spot position on the sample, derived from photoluminescence. Experimental data: lower (LP, empty dots) and middle polariton (MP, solid dots) branches. The full lines are a fit to the experiments with a three-coupled-mode model, using the plotted bare cavity-photon ($f$, dashed-dotted) and exciton ($X_{1s}$ and $X_{2s}$, dashed) curves, and Rabi-splitting of $\Omega _{f-1s}\simeq 13\,$meV and $\Omega _{f-1s}\simeq 3.5\,$meV, as input. \textbf{(b)} Photon ($|A_{f}|^{2}$, full line) and exciton ($|A_{1s}|^{2}$ and $|A_{2s}|^{2}$, dashed and dash-dotted lines) strengths of the MP, derived from the fit in panel (a). $A_{1s2s}^{2}=A_{1s}A_{2s}^{\ast }+A_{1s}^{\ast }A_{2s}$ (short-dashes), is the exciton crossed interference term.}
\label{fig1}
\end{figure}
%%%%%%%%%%%%%%%%%%%%%%%%%%%%%%%%%%%%%%%%%%%%%%%%%%%%%%%%%%%%%%%%%%%%%%%%%%%%%%%
The dispersion of the middle and lower polariton branches (MP and LP, respectively), derived from photoluminescence experiments, is displayed in Fig.~\ref{fig1}a as function of the spot position $x$ (in mm) on the sample. Unfortunately no direct observation of the presence of the upper polariton (UP) branch was found with luminescence. Using a simple three-coupled-mode model, the best fits of the experimental MP and LP branches was obtained for a Rabi-splitting of $\Omega _{f-1s}\simeq 13\,$meV and $\Omega _{f-2s}\simeq 3.5\,$meV between the cavity mode and the $1s$ and $2s$ excitons, respectively. The dashed and dash-dotted lines correspond to the bare (non-interacting) exciton ($X_{1s}$ and $X_{2s}$), and cavity-photon ($f$) dispersions used in the model, respectively.\cite{Bruchhausen-PRB68-205326(03)} The vertical dotted lines indicate the positions of the anticrossings, i.e. the minimal distance between the MP-LP branches and the UP-MP branches. The slight positive slope of the bare exciton dispersions is due to the small thickness gradient, that also affects the QW's.

Two different and independent Raman experiments were performed observing the inelastic scattered photons due to the Raman allowed CdTe QW longitudinal optical (LO) phonons. These phonons have an energy of $\sim 172\,$cm$^{-1}$ ($\sim 21.3\,$meV). The first experiment, consisted in setting the Stokes-shifted photons always at exact outgoing resonance with the MP, and changing the cavity-photon--exciton detuning by displacing the laser spot on the sample, as described in Refs.~[\onlinecite{Fainstein-PRB57-R9439(98), Bruchhausen-PRB68-205326(03)}]. The second experiment consisted in keeping the detuning fixed (i.e. at a fixed spot position), and varying the laser energy at near outgoing resonance, in order to scan the MP with the Stokes-shifted photons. In both cases the spectra were analyzed using a Jobin-Yvon T64000 triple spectrometer in substractive mode, equipped with a liquid-N$_{2}$-cooled charged-coupled device (CCD).

%%%%%%%%%%%%%%%%%%%%%%%%%%%%%%%%%%%%%%%%%%%%%%%%%%%%%%%%%%%%%%%%%%%%%%%%%%%%%%%
%%%%%%%%%%%%%%%%%%%%%%%%%%%%%%%%%%%%%%%%%%%%%%%%%%%%%%%%%%%%%%%%%%%%%%%%%%%%%%%
\section{Experimental Results}\label{Experimental Results} 
%%%%%%%%%%%%%%%%%%%%%%%%%%%%%%%%%%%%%%%%%%%%%%%%%%%%%%%%%%%%%%%%%%%%%%%%%%%%%%%
%%%%%%%%%%%%%%%%%%%%%%%%%%%%%%%%%%%%%%%%%%%%%%%%%%%%%%%%%%%%%%%%%%%%%%%%%%%%%%%

%%%%%%%%%%%%%%%%%%%%%%%%%%%%%%%%%%%%%%%%%%%%%%%%%%%%%%%%%%%%%%%%%%%%%%%%%%%%%%%
\begin{figure}[ttt]
\begin{center}
\includegraphics*[keepaspectratio=true, clip=true, angle=0, width=0.9\columnwidth]{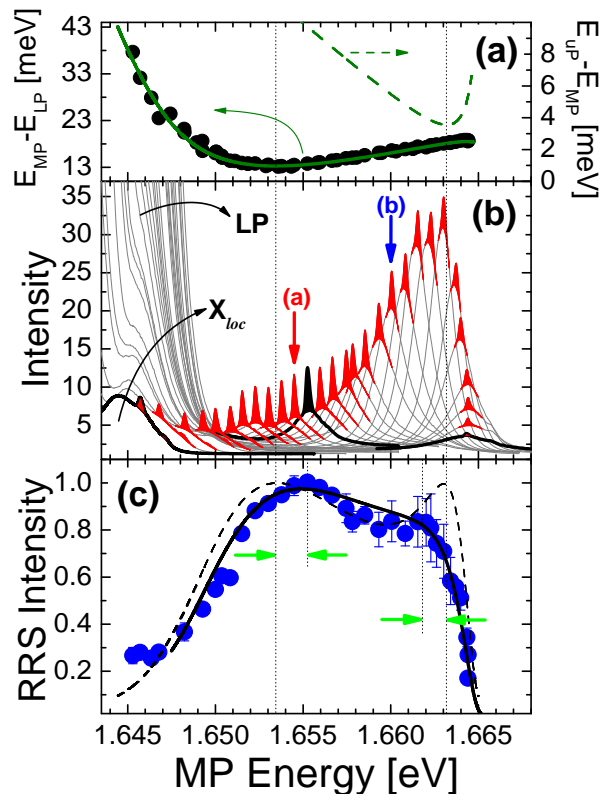}
\end{center}
\caption{(Color online) \textbf{(a)} Experimental and fitted differences between the energies of the MP and LP branches (left axis), and UP and MP (right axis), as function of the MP energy derived form Fig.~\ref{fig1}a. \textbf{(b)} Raman spectra at exact outgoing resonance with the MP for varying detuning. The Raman peaks are indicated by the filled areas. The arrows labeled as (a) and (b) indicate the equivalent positions of the Raman scans of Fig.~\ref{fig3}a and \ref{fig3}b respectively. \textbf{(c)} Resonant Raman intensity (full dots) derived from panel (b). The dashed curve corresponds to the calculation using the simple model without damping effects, and the full curve corresponds to the fit using the theory derived in the text. The vertical dotted lines indicate the position of both anticrossings. And the horizontal arrows in panel (c) indicate the shift of the RRS intensity maxima.}
\label{fig2}
\end{figure}
%%%%%%%%%%%%%%%%%%%%%%%%%%%%%%%%%%%%%%%%%%%%%%%%%%%%%%%%%%%%%%%%%%%%%%%%%%%%%%%
Typical spectra for the RRS experiment (of the first kind described in section \ref{Experimental Details}), are shown in Fig.~\ref{fig2}b, for different MP mode energies (i.e., laser spot positions). The Raman peak is highlighted on top of the MP luminescence. For each spectrum the exact resonance was set tuning the laser energy in order to maximize the Raman peaks intensity.
 
In the bottom panel (Fig.~\ref{fig2}c) the Raman intensity, derived from the spectra in panel (b), is shown as function of the MP energy. As additional information in the top panel (Fig.~\ref{fig2}a) the difference between the measured MP and LP energies is also displayed. The full (dashed) line corresponds to the calculated differences $E_{MP}-E_{LP}$ ($E_{UP}-E_{MP}$), using the fitted curves and parameters of Fig.~\ref{fig1}a obtained from the three-coupled-mode model. The vertical dotted lines indicate the anticrossing positions, i.e. the minima of the respective energy differences.
 
In Fig.~\ref{fig2}b, for low MP energies, luminescence corresponding to localized excitonic states (labeled as $X_{loc}$) can be distinguished.\cite{Fainstein-PRB57-R9439(98)} And for increasing MP energies, i.e. for increasing photon--$X_{1s}$ detuning, the overwhelming LP luminescence appears from the left. One thing to notice is how the MP luminescence increases when its energy approaches the $X_{2s}$ exciton at $\sim 1.665\,$eV. Note that the luminescence maximum is reached slightly below this energy. This typical behavior of luminescence in the presence of an anticrossing is refered to as ``cavity pulling'', in this case between the MP and the UP.\cite{Stanley-PRB53-10995(96)} A similar behavior is observed for the luminescence coming from the LP branch, slightly before the anticrossing of the MC photon and the $X_{1s}$ exciton.

%%%%%%%%%%%%%%%%%%%%%%%%%%%%%%%%%%%%%%%%%%%%%%%%%%%%%%%%%%%%%%%%%%%%%%%%%%%%%%%
\begin{figure}[t]
\begin{center}
\includegraphics*[keepaspectratio=true, clip=true, angle=0, width=0.7\columnwidth]{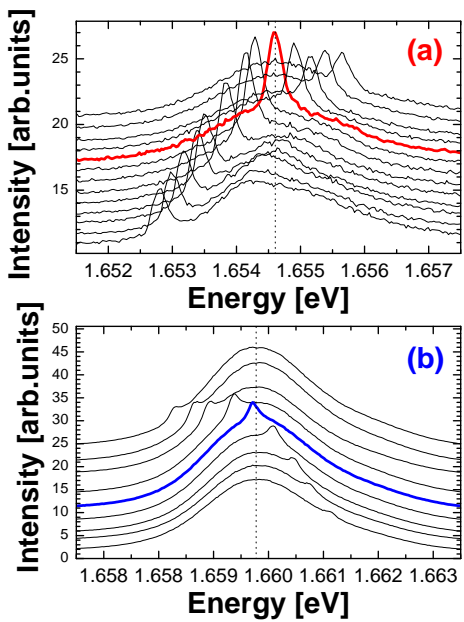}
\end{center}
\caption{(Color online) Outgoing Raman scans for a fixed photon-exciton detuning, and varying laser excitation energy. Scan (a) corresponds to a spot position of $x\simeq 2.5\,$mm and (b) to $x\simeq 3.1\,$mm. The Raman peak moves on top of the broad MP luminescence peak and resonates at its position, indicated by the vertical dotted line. The resonant situation is highlighted with a thicker curve, and corresponds to the positions indicated with vertical arrows in Fig.~\ref{fig2}, and labeled with (a) and (b) respectively. Note that the broad luminescence peak in (b) is $\sim 3$ times larger than in (a).}
\label{fig3}
\end{figure}
%%%%%%%%%%%%%%%%%%%%%%%%%%%%%%%%%%%%%%%%%%%%%%%%%%%%%%%%%%%%%%%%%%%%%%%%%%%%%%%
As for the second type of experiment, typical first order LO-phonon Raman scans for varying laser energy close to outgoing resonance with the MP at two different spot positions (i.e., detunings) are shown in Fig.~\ref{fig3}. In Fig.~\ref{fig3}a the MP is centered at $\sim 1.6548\,$eV (corresponding to a spot position on the sample of about $x\simeq 2.5\,$mm), whereas in Fig.~\ref{fig3}b it is centered at $\sim 1.6599\,$eV (for $x\simeq 3.1\,$mm). The spectra corresponding to the outgoing MP resonant condition are displayed with thicker lines, and the corresponding equivalent resonant spectra are indicated with arrows in Fig.~\ref{fig2}b. Note that the photoluminescence background of both scans in Fig.~\ref{fig3} differ considerably, as is clearly observed in the corresponding resonant spectra on Fig.~\ref{fig2}b. Figure \ref{fig4} shows the Raman intensity profiles (a) and (b) derived respectively from Fig.~\ref{fig3}a and \ref{fig3}b. A third profile (c) is also displayed, which corresponds to an energy of the MP at $\sim 1.6533\,$eV (for a spot position $x\simeq 2.2\,$mm),\footnote{This scan was shown in Ref.\onlinecite{Bruchhausen-PRB68-205326(03)}. The temperature was mistakenly indicated as $2.3\,$K, instead of $4.5\,$K.}.
Note the asymmetry of the intensity scans labeled as (a) and (c) in Fig.~\ref{fig4} towards lower energies. This might be ascribed to the strong presence of the LP. At lower energies the resonant contribution due to the LP starts to contribute to the Raman peaks intensity, thus inducing the asymmetry in the Raman scan. Scans (a) and (c) are taken near the anticrossing, where both branches (MP and LP) are closer to each other. This is not the case for the scan (b) (see Fig.~\ref{fig3}b and \ref{fig4}). This scan is far from the LP and at this detuning its resonant contribution is less strong. Thus this intensity scan is more symmetric. 

In the next section the theoretical background will be addressed, the
Green's function formalism presented, and the calculation of the Raman
intensity within this model will be described.

%%%%%%%%%%%%%%%%%%%%%%%%%%%%%%%%%%%%%%%%%%%%%%%%%%%%%%%%%%%%%%%%%%%%%%%%%%%%%%%%
%%%%%%%%%%%%%%%%%%%%%%%%%%%%%%%%%%%%%%%%%%%%%%%%%%%%%%%%%%%%%%%%%%%%%%%%%%%%%%%%
\section{Theoretical description}\label{Theory} 
%%%%%%%%%%%%%%%%%%%%%%%%%%%%%%%%%%%%%%%%%%%%%%%%%%%%%%%%%%%%%%%%%%%%%%%%%%%%%%%%
%%%%%%%%%%%%%%%%%%%%%%%%%%%%%%%%%%%%%%%%%%%%%%%%%%%%%%%%%%%%%%%%%%%%%%%%%%%%%%%%

%%%%%%%%%%%%%%%%%%%%%%%%%%%%%%%%%%%%%%%%%%%%%%%%%%%%%%%%%%%%%%%%%%%%%%%%%%%%%%%%
\subsection{The model}\label{Sec model}
%%%%%%%%%%%%%%%%%%%%%%%%%%%%%%%%%%%%%%%%%%%%%%%%%%%%%%%%%%%%%%%%%%%%%%%%%%%%%%%%

%%%%%%%%%%%%%%%%%%%%%%%%%%%%%%%%%%%%%%%%%%%%%%%%%%%%%%%%%%%%%%%%%%%%%%%%%%%%%%%
\begin{figure}[ttt]
\begin{center}
\includegraphics*[keepaspectratio=true, clip=true, angle=270, width=0.7\columnwidth]{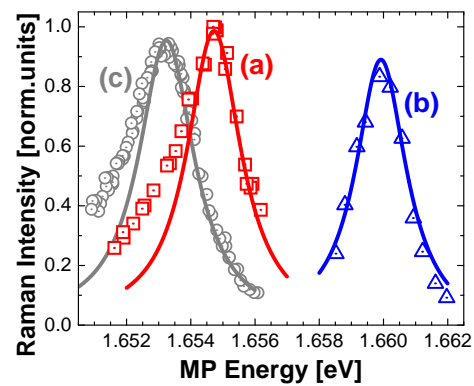}
\end{center}
\caption{(Color online) Raman intensity scans for varying laser energy close to the MP outgoing resonance for three different detuning. The MP energy corresponds respectively to (a) $E_{MP}\sim 1.6548\,$eV, (b) $\sim 1.6599\,$eV, and (c) $\sim 1.6533\,$eV. And the positions on the sample are (a) $x\simeq 2.5\,$mm, (b) $\simeq 3.1\,$mm, and (c) $\simeq 2.2\,$mm. The Raman intensity resonates at the MP energy. The experimental intensity labeled as (a) and (b) are derived from the spectra shown in Fig.~\ref{fig3}a and \ref{fig3}b respectively. The full lines are a fit to the data using the theory derived in the text (Sec.~\ref{Theory}).}
\label{fig4}
\end{figure}
%%%%%%%%%%%%%%%%%%%%%%%%%%%%%%%%%%%%%%%%%%%%%%%%%%%%%%%%%%%%%%%%%%%%%%%%%%%%%%%

In a cavity polariton system, only MC confined photons and QW confine excitons with the same in-plane wave vector ($k_{\parallel}$) are coupled.\cite{Jorda-PRB48-1669(93)} Since the above described experiments are performed in near-backscattering geometry, i.e. $z(x^{\prime},x^{\prime})\bar{z}$, $k_{\parallel}$ is considered to be zero ($k_{\parallel}= 0$).\footnote{For this polarization the electric field couples to excitonic states with $\Gamma_{5}$ symmetry of the $D_{2d}$ group,\cite{Jorda-PRB48-1669(93)} corresponding in this case to the heavy-hole excitons states $1s$ and $2s$.} In what follows the total wave vector and polarization indices will be suppressed, in order to simplify the notation.

The following Hamiltonian, for a given $k_{\parallel}$, is considered to describe the cavity exciton-polariton system:\cite{LeonHilario-JPCondMatter19-176210(07)} 
\begin{equation}\label{Hamiltonian}
H=H_{f}+H_{e}+H_{c}+H_{\text{mix}}\ .
\end{equation}
The first term corresponds to the photonic part of the Hamiltonian: 
\begin{equation}\label{Hamiltonian photon part}
H_{f}=E_{f}f^{\dagger }f+\sum_{p}\epsilon _{p}r_{p}^{\dagger}r_{p}+\sum_{p}(V_{p}r_{p}^{\dagger }f+\text{H.c.}).
\end{equation}
Here $f^{\dagger }$ creates a MC photon, which is hybridized to a continuum of radiative modes described by creation operators $r_{p}^{\dagger }$. The effect of the latter is to provide a finite life time ($\gamma_{f}$) to the cavity mode, even in the absence of light-matter interactions.

Similarly, the excitonic part of $H$ is described as 
\begin{equation}\label{Hamiltonian exciton part}
H_{e}=\sum_{i}E_{i}e_{i}^{\dagger }e_{i}+\sum_{iq}\epsilon
_{iq}d_{iq}^{\dagger }d_{iq}+\sum_{iq}(V_{iq}d_{iq}^{\dagger }e_{i}+\text{%
H.c.})\ ,
\end{equation}
where $e_{i}^{\dagger }$ creates exciton states that will couple strongly to the cavity photon mode. For example $e_{1}^{\dagger }$ creates a $1s$ exciton state of energy $E_{1s}$. Analogous to the case of the MC photon, we assume that each exciton mode mixes with a continuum of bosonic excitations (described by the operators $d_{ig}^{\dagger}$), which broadens its spectral density. The effect of this mixing is to introduce a broadening $\gamma_{ei}$ for the excitons even without interaction with the light. 
The detailed description of this broadening is beyond the scope of this work. Numerous physical mechanisms are known to contribute to the damping of the excitonic wave function and hereby to the line broadening. Some of them are for example: impurities, defect-scattering, interface roughness, well-width fluctuations, and alloy disorder, which have been demonstrated to be present in this kind of II-VI based microcavities.\cite{Bruchhausen-PRB73-85305(06)}
Whittaker has shown that the inhomogeneous broadening introduced by disorder can be described accurately introducing an imaginary part in the self energy of the exciton Green function.\cite{Whittacker-PRL77-4792(96),Whittacker-PRL80-4791(98)}
As we show below, $H_{e}$ has a similar effect within our theory. Therefore the broadenings $\gamma _{ei}$ [see Eq.~(\ref{deltas})] can be regarded as due to lifetime broadening.

The third term in Eq.~(\ref{Hamiltonian}) describes the electronic excitations that do \emph{not} couple strongly to the MC photon mode, 
\begin{equation}\label{Hamiltonian el-hole excitations}
H_{c}=\sum_{k}\epsilon _{k}a_{k}^{\dagger }a_{k}\ .
\end{equation}

These states include discrete exciton states as well as states of the exciton-continuum (i.e. the continuum of electron-hole excitations). The latter correspond to the eigenstates of the exciton hydrogen-like Hamiltonian that are not bound and have positive internal energy.\cite{Shinada-JPhysSocJap21-1936(66), Chao-PRB43-6530(91)} The operators, which describe this exciton continuum, can be well approximated by $a_{k}^{\dagger }=c_{k}^{\dagger }v_{k}$, where $v_{k}$ destroys a valence electron with 2D wave vector $k$ and polarization $\sigma$, and $c_{k}^{\dagger}$ creates an electron in the conduction band with the same $k$ and $\sigma$ (so that the total wave vector $k_{\parallel}=0$). Note that the commutator $[a_{k},a_{k}^{\dagger}]=v_{k}^{\dagger}v_{k}-c_{k}^{\dagger} c_{k}$. For the temperatures and laser intensities of the experiments, all valence (conduction) states are practically occupied (unoccupied) and therefore with high accuracy $v_{k}^{\dagger }v_{k}=1$ ($c_{k}^{\dagger }c_{k}=0$). Therefore the operators entering Eq. (\ref{Hamiltonian el-hole excitations}) can be regarded as ordinary bosons.\cite{Mahan-ManyParticlePhysics(00), Cho-Excitons(79)}

The separation of the bound excitonic states into those included in $H_{e}$ or $H_{c}$ is in principle arbitrary. In particular, when the width of the weakly bound excitonic states near the bottom of the gap is larger than the separation in energy between two next states, these states can be well described as a continuum, beginning at some energy $E'_{X_{c}}$, slightly below the energy of the first excitonic state not included in $H_{e}$.
In our specific treatment for the experimentally studied sample, we
describe all states included in $H_{c}$ as a continuum beginning at $E'_{X_{c}}$,
and include the two lowest excitonic states ($E_{1s}$ and $E_{2s}$) that couple ``strongly'' to the MC photon in $H_{e}$. For simplicity we take $E'_{X_{c}}=E_{2s}$.

Finally, the last term of the Hamiltonian Eq. (\ref{Hamiltonian}) describes the mixing between light and matter 
\begin{equation}\label{Hamiltonian light-matter}
H_{\text{mix}}=\sum_{i}(V_{i}e_{i}^{\dagger }f+\text{H.c.}%
)+\sum_{k}(V_{k}a_{k}^{\dagger }f+\text{H.c.}).
\end{equation}
According to the quantum theory of radiation, in the dipolar approximation $V_{k}$ is proportional to $\langle k_{c}|p_{\sigma }|k_{v}\rangle$, where $p_{\sigma }$ is the momentum operator in the direction of the electric field, and $|k_{c}\rangle$ ($|k_{v}\rangle)$ is the wave function which corresponds to the operator $c_{k}$ ($v_{k}$). For simplicity, we are neglecting terms which do not conserve the number of bosons. They can be included without much complications, but their effect is small for the energies of interest.\cite{Jorda-PRB50-2283(94)}\\

The Hamiltonian Eq. (\ref{Hamiltonian}) is quadratic and can be formally diagonalized by a Bogoliubov transformation to the form
\begin{equation}\label{Hamiltonian diagonal}
H=\sum_{\nu} E_{\nu } p_{\nu }^{\dagger }p_{\nu }\ ,
\end{equation}
where the generalized polariton operators $p^{\dagger}_{\nu}$, are related to the creation operators entering Eq. (\ref{Hamiltonian}) by the linear combination 
\begin{equation}\label{polariton operators}
p_{\nu }^{\dagger }=\sum_{j}A_{\nu j}\beta_{j}^{\dagger}\ .
\end{equation}
$\beta_{j}^{\dagger}$ stands for any operator entering Eq.~(\ref{Hamiltonian}), and $A_{\nu j}$ are the coefficients that give the (detuning dependent) corresponding photonic or excitonic weight of the polariton state $\nu$.\\

Note that if only the first terms in Eqs.~(\ref{Hamiltonian photon part}), (\ref{Hamiltonian exciton part}) and (\ref{Hamiltonian light-matter}) are kept, the simple standard coupled-mode model is obtained.\cite{Bruchhausen-PRB68-205326(03), Bruchhausen-AIPCP772-1117(05)} For instance, the three-coupled-mode model results, when the MC mode couples strongly only to the $i=1s$ and the $i=2s$ exciton states, and the values of the interaction are given by $2\,V_{1}\equiv\Omega_{f-1s}$ and $2\,V_{2}\equiv\Omega_{f-2s}$, as it is the case for the above described experiments.

%%%%%%%%%%%%%%%%%%%%%%%%%%%%%%%%%%%%%%%%%%%%%%%%%%%%%%%%%%%%%%%%%%%%%%%%%%%%%%%%
\subsection{The Raman intensity}\label{Sec Raman Intensity} 
%%%%%%%%%%%%%%%%%%%%%%%%%%%%%%%%%%%%%%%%%%%%%%%%%%%%%%%%%%%%%%%%%%%%%%%%%%%%%%%%

It has been shown previously,\cite{Fainstein-PRB57-R9439(98), Bruchhausen-PRB68-205326(03), Bruchhausen-AIPCP772-1117(05), LeonHilario-JPCondMatter19-176210(07), LeonHilario-InPress-(07)} that the outgoing Raman cross section mediated by cavity polaritons can be calculated by first order perturbation theory as 
\begin{equation}  \label{scattered Raman intensity}
I(\omega)\propto T_{i}\,W_{i\rightarrow s}\,T_{s}\ ,
\end{equation}
where $T_{i}$ describes the incoming channel, i.e. the conversion of the incident external photon to the initial polariton state inside the sample. For the case of outgoing resonance and the scattering geometry used in the experiments, $T_{i}$ is essentially proportional to the DBR's residual transmission, and can be taken as constant. $W_{i\rightarrow s}$ is the scattering probability from the initial $|i\rangle$ to the final $|s\rangle$ polariton state per unit of time. $T_{s}$ describes the transmission of the final polariton states to the final photon state outside the sample, which is detected. Since the final polariton state leaves the sample due to its photonic part, $T_{s}$ is essentially the projection of the polariton state on the final photon continuum state, and is thus proportional to the mediating polaritons photon strength, given by $|A_{\nu f}|^{2}$.

We will center our attention on $W_{i\rightarrow s}$, which actually describes the \emph{inelastic scattering process} itself. The transition probability is given by Fermi's golden rule as 
\begin{equation}\label{tramsition probability}
W_{i\rightarrow s}=\frac{2\pi }{\hbar }|\langle i|H'|s\rangle
|^{2}\rho(\omega),
\end{equation}
where the initial polariton state is $|i\rangle=p^{\dagger}_{\nu^{\prime}}\,|0\rangle$, the final state is given by $|s\rangle= p^{\dagger}_{\nu}\, b^{\dagger}\,|0\rangle$, where $b^{\dagger}$ creates a LO phonon, and $H'$ is the interaction between polaritons and LO phonons (e.g. Fr\"ohlich-interaction) through the excitonic part of the former. And $\rho(\omega)$ stands for the density of final polariton states and is given by
\begin{equation}\label{total density of states}
\rho(\omega )=\sum_{\nu}\delta(\omega-E_{\nu})\ .
\end{equation}

Since the Hilbert space defined by (\ref{Hamiltonian diagonal}) and (\ref{polariton operators}) is infinite, it is cumbersome to work with the eigenstates and eigenenergies. 
However, we do not need them to obtain the intensity of the Raman scattering process. It is enough to use retarded Green's functions, that involve the photon and exciton operators ($\beta_{l}^{\dagger}$). These functions $G_{jl}(\omega)=\langle \langle \beta_{j};\beta_{l}^{\dagger}\rangle \rangle _{\omega}$ can be obtained solving a system of equations derived from the equations of motion for the Green's function: 
\begin{equation}\label{Eqn of motion for Green fkt}
\omega \langle \langle \beta_{j};\beta_{l}^{\dagger }\rangle \rangle
_{\omega}=\delta _{jl}+\langle \langle \lbrack
\beta_{j},H];\beta_{l}^{\dagger }\rangle\rangle _{\omega}\ .
\end{equation}

The polariton-phonon interaction $H'$ is proportional to the exciton part of the scattered polariton. Thus, for the case of a system where only two exciton states interact strongly with the MC mode, the Raman intensity of Eq.~(\ref{scattered Raman intensity}) results proportional to 
\begin{equation}\label{scattered Raman intensity 2}
I(\omega)\propto |A_{\nu f}|^{2}\,|A_{\nu e1}+\alpha A_{\nu e2}|^{2}\,\rho(\omega)\ ,
\end{equation}
where $\nu$ is the label of the polariton eigenstate such as $E_{\nu}=\omega$.

Here we are neglecting the contribution of the electron-hole continuum to $%
H'$, and $\alpha $ represents the ratio of matrix elements of the exciton-LO phonon interaction between $2s$ and $1s$ excitons. Note that if lifetime effects are neglected, i.e. if all states but the three polariton branches are left aside, $\rho _{\nu}(\omega)$ is reduced to $\sum_{\nu}\delta (\omega-E_{\nu})$, for $\nu=LP$, $MP$, and $UP$, and the simple three-mode model for the RRS intensity is recovered.\cite{Bruchhausen-PRB68-205326(03)}

Using the Lehman representation of the Green's function\cite{Mahan-ManyParticlePhysics(00)} it can be shown that 
\begin{eqnarray}
\rho _{jl}(\omega ) &=&-\frac{1}{2\pi }\big[G_{jl}(\omega
+i0^{+})-G_{jl}(\omega -i0^{-})\big]  \nonumber  \label{state density matrix}
\\
&=&A_{\nu j}\bar{A}_{\nu l}\rho(\omega )\ ,
\end{eqnarray}
and taking into account that $\sum_{j}|A_{\nu j}|^{2}=1$, and that $\rho(\omega)=\sum_{j}\rho _{jj}(\omega)$, it can be followed by replacing these expressions in Eq.~(\ref{scattered Raman intensity 2}), that the desired equation for the Raman intensity is given by 
\begin{equation}
I(\omega )\propto \frac{\rho _{ff}[\rho _{e1,e1}+|\alpha|^{2}\rho _{e2,e2}+2\text{Re}(\alpha \,\rho _{e2,e1})]}{\sum_{j}\rho _{jj}}\ .
\label{scattered Raman intensity 3}
\end{equation}
In the denominator we neglect the contribution of the continuum states, because it is negligible near the resonance condition for the outgoing polariton. Therefore the sum over $j$ includes only the densities of excitons and photons. In particular if only two excitons are involved in the polariton system, we need to calculate only the diagonal Green's functions $G_{ff}(\omega )$, $G_{e1,e1}(\omega)$, and $G_{e2,e2}(\omega)$. And also the crossed Green's function $G_{e2,e1}(\omega)$. Using the equations of motion we obtain: 
\begin{eqnarray}
G_{ff}(\omega ) &=&\frac{1}{\tilde{\omega}_{f}-\sum_{i}\frac{V_{i}^{2}}{\tilde{\omega}_{i}}-S_{f}^{\prime }}\ ,  \nonumber  \label{Green fkts} \\
G_{e1,e1}(\omega ) &=&\frac{1}{\tilde{\omega}_{1}-\frac{V_{1}^{2}}{\tilde{\omega}_{f}-V_{2}^{2}/\tilde{\omega}_{2}}}\ ,  \nonumber \\
G_{e2,e2}(\omega ) &=&\frac{1}{\tilde{\omega}_{2}-\frac{V_{2}^{2}}{\tilde{\omega}_{f}-V_{1}^{2}/\tilde{\omega}_{1}}}\ ,  \nonumber \\
G_{e2,e1}(\omega ) &=&G_{e1,e2}(\omega )  \nonumber \\
&=&\frac{V_{1}V_{2}}{\tilde{\omega}_{1}\tilde{\omega}_{2}\tilde{\omega}_{f}-V_{1}^{2}\tilde{\omega}_{2}-V_{2}^{2}\tilde{\omega}_{1}}\ ,
\end{eqnarray}
where 
\begin{equation}
\tilde{\omega}_{f}=\omega -E_{f}-S_{f}\;\;,\;\;\;\;\;\tilde{\omega}%
_{i}=\omega -E_{ei}-S_{ei}\ ,  \label{wtilde}
\end{equation}%
and 
\begin{eqnarray}
S_{f}(\omega ) &=&\sum_{p}\frac{|V_{p}|^{2}}{\omega +i0^{+}-\epsilon _{p}}\ ,\nonumber  \label{sums} \\
S_{ei}(\omega ) &=&\sum_{q}\frac{|V_{iq}|^{2}}{\omega +i0^{+}-\epsilon _{iq}}\ ,  \nonumber \\
S_{f}^{\prime }(\omega ) &=&\sum_{k}\frac{|V_{k}|^{2}}{\omega+i0^{+}-\epsilon _{k}}\ .
\end{eqnarray}%
For the first two of these sums, we assume that the results are imaginary constants that we take as parameters: 
\begin{equation}
S_{f}(\omega )=-i\gamma _{f}\ ,\qquad S_{j}(\omega )=-i\gamma _{ej}\ .
\label{deltas}
\end{equation}
This is the result expected for constant density of states and matrix elements. These two approximations are clearly not valid for the last sum. The last sum in Eq~(\ref{sums}), corresponds to the excitons that do not interact strongly with the MC photons, and to the electron-hole continuum. This continuum begins at the energy of the gap and corresponds to vertical transitions in which the light promotes a valence electron with 2D wave vector $k_{\parallel}$ to the conduction band with the same wave vector. In the effective-mass approximation, the energy $\epsilon_{k}$ is quadratic with $k_{\parallel}$ and this leads to a constant density of states beginning at the gap. As mentioned in the previous section, $V_{k}\sim\langle k_{c}|p_{\sigma}|k_{v}\rangle$. Approximating the wave functions as plane waves and averaging over all directions of $k_{\parallel }$ one has $\langle |V_{k}|^{2}\rangle \sim k_{\parallel }^{2}/2$, proportional to the energy $\epsilon_{k}$. Using these assumptions and separating $S_{f}^{\prime }(\omega )$ in real [$r(\omega)$] and imaginary [$y(\omega)$] parts we can write: 
\begin{eqnarray}
S_{f}^{\prime }(\omega ) &=&r(\omega )-i\,y(\omega ),\text{ }r(\omega )=\frac{1}{\pi }\int d\epsilon \frac{y(\epsilon )}{(\omega -\epsilon )}\ ,\label{re} \\
y(\omega ) &=&\mathcal{A}~(\omega -E_{Xc})\,\Theta (\omega -E_{Xc}),\label{im}
\end{eqnarray}
where $E_{Xc}$ is the bottom of the electron-hole continuum, $\mathcal{A}$ is a dimensionless parameter that accounts for the coupling magnitude of the interaction between the MC photon and the continuum of excitations, and $\Theta (\omega)$ stands for the Heaviside function (step function). The real part $r(\omega)$ can be absorbed in a renormalization of the photon energy and is unimportant in what follows. The imaginary part $y(\omega)$ is a correction to the photon width for energies above the bottom of the continuum. From the formalism outlined above, this is expected to lead to a decrease in the RRS intensity when the energy is above the bottom of the continuum. The expression for $y(\omega)$ accounts so far only for effects of the exciton-continuum, and leads therefore to a discontinuous first derivative of the calculated RRS profile. This discontinuity is overpassed, if the remaining bound exciton states are considered. To account for this infinite number of bound states, which become very dense below the bottom of the continuum, the lower part of $y(\omega)$ is replaced by a parabolic
function that matches $(\omega-E_{Xc})$ with a continuous derivative, starting from zero at the energy ($E'_{X_{c}}$). At the energy $E'_{X_{c}}$, which corresponds to the bottom of the electronic excitations in $H_{c}$, as described in the previous subsections:
\begin{equation}
y(\omega )=\,\left\{ 
\begin{array}{l l}
0 & \text{if } \omega\leq E'_{X_{c}} \\ 
\mathcal{A}~\frac{(\omega-E'_{X_{c}})^{2}}{4(E_{Xc}-E'_{X_{c}})} & \text{if } E'_{X_{c}}\leq\omega\leq 2E_{Xc}-E'_{X_{c}} \\ 
\mathcal{A}~(\omega-E_{Xc}) & \text{if }\omega \geq 2E_{Xc}-E'_{X_{c}}
\end{array} 
\right.\ . \label{im2}
\end{equation}

As it will be show in the next section, this refinement in the description of the continuum electron-hole electronic excitations and weakly bound excitons does not affect the Raman profile in resonance with the LP and MP branches, and only affects the resonant profile on the UP branch for energies above $E'_{X_{c}}$, leading to smoother curves.\\

The Raman intensity $I(\omega)$ for a fixed detuning is given by Eq.~(\ref{scattered Raman intensity 3}), together with Eqs.~(\ref{state density matrix}), (\ref{Green fkts})--(\ref{deltas}),(\ref{im2}) and its dependence with detuning (or equivalently with the spot position) enters through the corresponding variation of the bare MC photon ($E_{f}$) and exciton ($E_{ei}$) energies. In the next section we will discuss this model, and compare the results to the above presented experimental data.

%%%%%%%%%%%%%%%%%%%%%%%%%%%%%%%%%%%%%%%%%%%%%%%%%%%%%%%%%%%%%%%%%%%%%%%%%%%%%%%%
%%%%%%%%%%%%%%%%%%%%%%%%%%%%%%%%%%%%%%%%%%%%%%%%%%%%%%%%%%%%%%%%%%%%%%%%%%%%%%%%
\section{Analysis and discussion}\label{Analisys and discussion} 
%%%%%%%%%%%%%%%%%%%%%%%%%%%%%%%%%%%%%%%%%%%%%%%%%%%%%%%%%%%%%%%%%%%%%%%%%%%%%%%%
%%%%%%%%%%%%%%%%%%%%%%%%%%%%%%%%%%%%%%%%%%%%%%%%%%%%%%%%%%%%%%%%%%%%%%%%%%%%%%%%

The role of the cavity polaritons as intermediate states in the scattering process is fully displayed in the resonant enhancement of the Raman intensity as function of exciton-photon detuning, as described in Sec.~\ref{Experimental Results} and shown in Fig.~\ref{fig2}. Here the intensity of the scattered resonant Raman signal tuned to a polariton mode (the MP in this case) is followed continuously as this mode changes its character from very excitonic ($1s$-type) to photonic, and again very excitonic ($2s$-type).

In a simple picture, since the interaction of the polariton with the LO-phonons is produced through the excitonic part of the polariton, the inelastic scattering process is favored when the polariton excitonic part is large. On the other hand the Stokes shifted polariton is only detected when the polariton is transmitted outside of the sample. Since this coupling to the photon-continuum is provided through the photonic part of the polariton, the Raman process is favored when this weight is also large. Thus in view of these two considerations, the process is optimized when a compromise between both situations is found, and this is reached at zero detuning, i.e. at the crossing of the bare exciton and MC photon, or what is equivalent at the energies where the differences $E_{MP}-E_{LP}$ or $E_{UP}-E_{MP}$ have a minimum. Thus, within this simplified picture, at these energies the RRS efficiency is maximized, and the scattered intensity decreases towards the pure excitonic and photonic limits. As we will see, the inclusion of photon and exciton lifetime broadening, i.e. the coupling of the MC photon to the exciton continuum described in Sec.~\ref{Theory}, introduces important modifications to this correct but simplified view.

For comparison, the results of calculating the RRS intensity using the simple model given by Eq.~(\ref{scattered Raman intensity 2}), considering only three branches (with no lifetime broadening) is shown in Fig.~\ref{fig2}c (dashed lines). Two maxima are displayed corresponding to the energy of the two anticrossings (compare to Fig~\ref{fig2}a). This can be understood when the corresponding strength\footnote{The strength is defined as the squared photonic or excitonic weight of the polariton, i.e. $|A_{\nu j}|^{2}$.} components of the MP are analyzed (plotted in Fig~\ref{fig1}b). The first anticrossing ($\sim 2.2\,$mm) happens at $E_{f}-E_{1s}=0$ where $|A_{f}|^{2}\simeq|A_{1s}|^{2}$ since the $2s$ strength at this position is very small. The second anticrossing ($\sim 3.4\,$mm) happens shortly before $E_{f}-E_{2s}=0$, where the total exciton part equals the photonic part. As explained above at these points the compromise is found, and a maximal scattering intensity is expected within this model. As one can see in Fig~\ref{fig2}c, this model qualitatively reproduces the general behavior well. However as pointed out before\cite {Fainstein-PRB57-R9439(98), Bruchhausen-PRB68-205326(03)} many features are clearly \emph{not} reproduced quantitatively. The first experimental maximum appears shifted ($\sim 2\,$meV) to higher energies, and something similar happens to the second peak, which appears shifted to lower energies ($\sim 1\,$meV) and less intense (shoulder like), as indicated with arrows in Fig.~\ref{fig2}c. These observations can be assigned to lifetime or damping effects,\cite{Bruchhausen-PRB68-205326(03)} and point to the need of the polariton mediated Raman theory described in the previous section (Sec.~\ref{Sec Raman Intensity}) that includes rigorously the lifetime broadening into the polariton states. In what follows, our experimental results will be compared to this theoretical model based on a Green's function approach.

The Raman intensity, is given by Eq.~(\ref{scattered Raman intensity 3}), together with Eqs.~(\ref{state density matrix}), (\ref{Green fkts})--(\ref{deltas}). Parameters $E_{f}$, $E_{1s}$, and $E_{2s}$ as function of the spot position $x$, as well as $2\,V_{1s}=\Omega_{f-1s}$ and $2\,V_{1s}=\Omega_{f-2s}$, were derived from the fit of the experimental dispersion shown in Fig~\ref{fig1}a, using the result of the three-coupled-mode model as a seed. We estimated the bottom of the exciton continuum energy at $E_{Xc}\simeq 1.670\,$meV. 
%para 2D en $x=0$ daria $E_{Xc}=1.666\,$eV, para 3D en $x=0$ daria $E_{Xc}=1.670\,$eV
Parameter $\mathcal{A}$ is taken as 0.031, derived from the fit to Raman experiments in resonance with the UP on a similar II-VI cavity in the very strong coupling regime.\cite{LeonHilario-InPress-(07)} The line broadening of the non interacting cavity-photon is estimated by standard realistic reflectivity calculations\cite{Yeh-JOptSocAm67-473(76),Yariv-IntroductionToOpticalElectronics(71)} to be $\gamma_{f}\simeq 1\,$meV. The exciton line broadening $\gamma_{e1}$, $\gamma_{e2}$, and the $2s$--$1s$ ratio of the exciton-LO phonon interaction matrix elements given by $\alpha$ (assumed as a real number), together with the proportionality factor, are left as fitting parameters. Both types of experiments described in Sec.~\ref{Experimental Results} (Fig~\ref{fig2}c and Fig~\ref{fig4}) are fitted \emph{simultaneously}. Summarizing in a simple interpretation, the first experiment for varying spot position changes the photon-exciton character of the mediating polariton. While for the second experiment, the resonance scan at a fixed spot position (fixed detuning), accounts for the \emph{total} polariton line
broadening for that particular exciton-photon mixture. 
%%%%%%%%%%%%%%%%%%%%%%%%%%%%%%%%%%%%%%%%%%%%%%%%%%%%%%%%%%%%%%%%%%%%%%%%%%%%%%%
\begin{figure}[ttt]
\begin{center}
\includegraphics*[keepaspectratio=true, clip=true, angle=270, width=0.8\columnwidth]{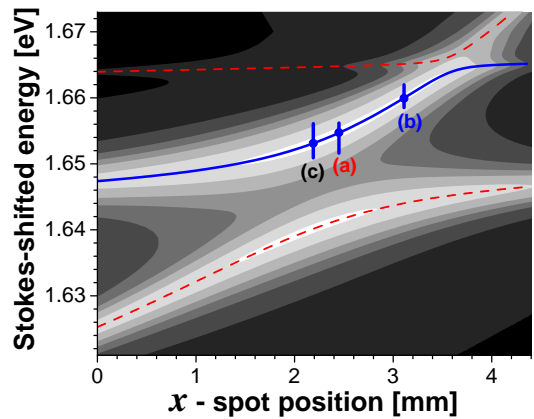}
\end{center}
\caption{(Color online) Contour plot of the total calculated outgoing Raman intensity. The LP and UP mode positions are indicated with dashed lines, and the MP outgoing resonant condition is plotted with a full curve. The Raman scans of Fig~\ref{fig4} are indicated with vertical segments and labeled (a), (b), and (c) respectively. The intensity is given in logarithmic scale.}
\label{fig5}
\end{figure}
%%%%%%%%%%%%%%%%%%%%%%%%%%%%%%%%%%%%%%%%%%%%%%%%%%%%%%%%%%%%%%%%%%%%%%%%%%%%%%%
Best fits were obtained for $\gamma_{e1}=1.2\,$meV and $\gamma_{e2}=0.9\,$meV, and $\alpha =-0.45$. The resulting curves are shown for the RRS in Fig.~\ref{fig2}c (full line) and for the resonant scans respectively in Fig.~\ref{fig4} (full lines). To fit the latter, only the upper half part of the scan has been used for the (b) and (c) scans, while for (a) the whole scan has been used. As can be observed the agreement with the measured data is very good, and provides a quantitative description. The obtained values for the parameters are relatively robust whether $\alpha$, $\gamma_{e1}$ or $\gamma_{e2}$ are changed. 

Since the $1s$- and $2s$-exciton weights for the three positions [scans (a),(c) and (b)] are different (see Fig.\ref{fig1}b), small changes in these parameters modify the width of the calculated lorentzian-type curves obtained for a fixed detuning, changing consequently also the shape of the RRS scan. The simultaneous fit of both independent experiments determines unambiguously the way the values that these parameters can adopt.

A point that might seem odd at first sight, is the fact that for the calculations $\gamma_{e1}$ results slightly larger than $\gamma_{e2}$.
Nevertheless, this result is to be expected if the origin of the broadening is disorder with a correlation length smaller or of the order of the 1s exciton's size:\cite{Whittacker-PRL77-4792(96), Whittacker-PRL80-4791(98)} the inhomogeneous broadening is substantially reduced by averaging over a volume larger than the length scale of the disorder potential. In fact, this is the reason why the disorder does not affect the broadening of the photons, the wave length of which is much larger than the disorder correlation length.\cite{Whittacker-PRL80-4791(98)}
%%%%%%%%%%%%%%%%%%%%%%%%%%%%%%%%%%%%%%%%%%%%%%%%%%%%%%%%%%%%%%%%%%%%%%%%%%%%%%%
\begin{figure}[ttt]
\begin{center}
\includegraphics*[keepaspectratio=true, clip=true, angle=0, width=0.95\columnwidth]{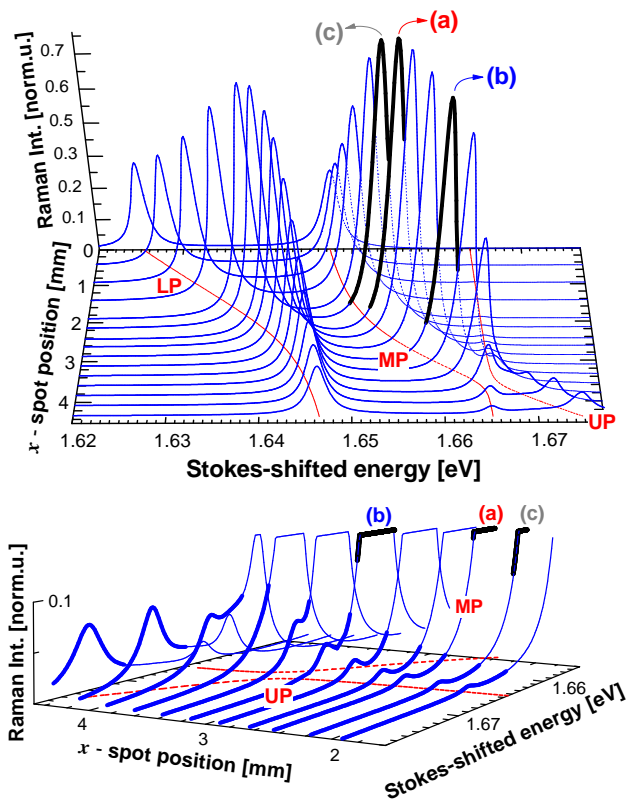}
\end{center}
\caption{(Color online) Three-dimensional plot of the calculated total outgoing Raman intensity profile as function of spot positions ($x$) and the corresponding Stokes-shifted energy. The full curves (LP, MP, and UP) projected on the $xy$-plane of the plot indicate the polariton dispersion of the polariton branches. The highlighted resonance profiles on the MP, marked as (a), (b), and (c) correspond respectively to the fitted curves in Fig.~\ref{fig4}. The bottom panel shows a \emph{rotated} detail of the UP region of the upper panel.}
\label{fig6}
\end{figure}
%%%%%%%%%%%%%%%%%%%%%%%%%%%%%%%%%%%%%%%%%%%%%%%%%%%%%%%%%%%%%%%%%%%%%%%%%%%%%%%

The total calculated Raman outgoing intensity is presented in Fig.~\ref{fig5} as a contour plot as function of the spot position ($x$) and the corresponding Stokes-shifted energy. The mode positions are indicated with dashed lines (LP and UP), and the MP outgoing resonant condition is plotted with a thick curve. In this figure the Raman intensity is shown in a logarithmic scale. The energies and $x$-positions of the Raman scans (shown in Fig.~\ref{fig4}) are also indicated with vertical segments [marked as (a), (b), and (c) respectively], and the corresponding exact resonant position with a full circle. 
%As can be seen in this figure, the non resonant Raman intensity between the LP and the MP modes increases slightly near the first anticrossing where the LP and MP modes are closer.
As can be seen in this figure the Raman intensity decreases slightly asymmetrically when moving away from the exact MP Raman outgoing resonant condition when the Stokes-shifted energy is varied near the position of the first anticrossing, where the LP and MP modes are closer. It has to be pointed out that the calculated asymmetry points in the right direction, but is unfortunately not enough to account for the stronger one observed experimentally for Raman profiles (a) and (c).
Something similar happens at the second anticrossing between the MP and UP, but in this case, due to the extremely low resonant outgoing Raman intensity on the UP, the anticrossing is not observable and is ``washed out''.

The same calculated total outgoing Raman intensity profile is presented in Fig.~\ref{fig6} as a three-dimensional plot. The mode positions (LP, MP) derived by projecting the corresponding resonance maxima on the $xy$-plane, are shown with dotted curves. 
For the case of the UP mode, as previously mentioned, it was not possible to determine the precise position of the RRS maximum of the UP at the second anticrossing. Thus, for this case, the curve shown corresponds to the one derived from the three-coupled-mode model (see Fig.~\ref{fig1}a), which differs very little from the one derived by projecting the corresponding UP resonance maxima (were the UP resonance maxima are clearly distinguishable).
%For the case of the upper polariton branch (UP) it was not possible to determine the precise position of the RRS maximum of the UP near the second anticrossing, and thus the curve shown corresponds to the one derived from the three-coupled-mode model (see Fig.~\ref{fig1}). 
The bottom panel of Fig.~\ref{fig6} shows a \textit{rotated} detail of the UP region. The UP resonant profiles are presented highlighted. Here the upper Raman resonant peak is barely seen. For small $x$, the UP appears as a tiny peak that increases slightly its intensity when approximating the second anticrossing, and merging shoulder like to the resonant profile coming from the MP, and increasing its energy after the anticrossing, decreasing its intensity.
As can be derived from Fig.~\ref{fig1}a, for small $x$ the UP has a strong excitonic character, which is conserved shortly before the second anticrossing. The small photonic component of the UP at these energies is very small (see Fig.~\ref{fig6}), and thus the coupling of the Stokes-shifted polaritons to the continuum of photonic states at the exterior of the cavity is very weak. When the position approaches the second anticrossing the photon-exciton mixture of the UP becomes larger, and therefore the Raman intensity increases, reaching a broad maximum slightly after this anticrossing. For increasing energies, the coupling of the electron-hole continuum to this mode becomes larger and consequently its life time is reduced, the RRS intensity decreases, and the Raman profile becomes significantly broader.

As can be seen from Fig.~\ref{fig6}, the scattered intensity for the outgoing resonant Raman condition with the UP is more than one order of magnitude smaller as compared to the situation in resonance with the MP. This extremely small light intensity scattered by the upper branch near the second anticrossing, together with the important intensity loss due to the interaction of the polariton with the exciton continuum (as explained above) are the most probable reasons that hindered the experimental observation of this mode. For completeness, the fitted Raman intensity profiles of Fig.~\ref{fig4} are shown on the same plot, also highlighted by thicker lines and respectively indicated as (a), (b), and (c).

In our theory, we have not considered the possibility of polaritons of a higher branch to be scattered to lower polariton states. Since the LP is the lowest excited state of the system, the inclusion of this effect would cause an important increase of the LP resonant outgoing Raman intensity, with respect to the resonant scattered intensity by the higher order polariton branches (MP, UP). This kind of effect has been reported by Tribe \textit{et al.}\cite{Tribe-PRB56-12429(97)} for a III-V type cavity with build-in electric field. The significantly larger contribution to the total scattered intensity coming from the LP, may explain the asymmetry of the observed Raman scans (see Fig.~\ref{fig4}), and would be particularly important at those situations were the LP and the MP are nearest, as is the case in for the Raman scans labeled (a) and (c) in the Figs.~\ref{fig4} to \ref{fig6}, i.e. near the first anticrossing.

A study of this processes, as well as detailed calculations of the different contributions to the exciton line width (supposed here as simple constants), and the careful determination of the exciton-LO phonon matrix elements (here accounted by the constant factor $\alpha$), might lead to refinements and improvements of the theory. 
However, we believe that the formal framework of how the main physical ingredients enter the polariton mediated Raman process, is well described by the above presented theory.

%%%%%%%%%%%%%%%%%%%%%%%%%%%%%%%%%%%%%%%%%%%%%%%%%%%%%%%%%%%%%%%%%%%%%%%%%%%%%%%%
%%%%%%%%%%%%%%%%%%%%%%%%%%%%%%%%%%%%%%%%%%%%%%%%%%%%%%%%%%%%%%%%%%%%%%%%%%%%%%%%
\section{Conclusions}\label{Conclusions} 
%%%%%%%%%%%%%%%%%%%%%%%%%%%%%%%%%%%%%%%%%%%%%%%%%%%%%%%%%%%%%%%%%%%%%%%%%%%%%%%%
%%%%%%%%%%%%%%%%%%%%%%%%%%%%%%%%%%%%%%%%%%%%%%%%%%%%%%%%%%%%%%%%%%%%%%%%%%%%%%%%

Summarizing, we have reported two types of independent Raman experiments in near resonance with the middle cavity-polariton branch of a II-VI based semiconductor optical cavity. The first experiment shows the behavior of the Raman intensity at exact outgoing resonance with the cavity-polariton, for varying photon-exciton detuning. And the second experiment shows the evolution for a fixed photon-exciton detuning, when the Stokes-shifted photons are tuned on and off resonance by varying the excitation energy. Both independent experiments are accounted \emph{simultaneously} by the here developed cavity-polariton mediated Raman theory, that rigorously includes photon and exciton damping effects. The presented theory describes quantitatively well the experimental results, a fact that remarks the importance of the exciton and cavity-photon lifetime effects in the process of inelastic scattering of light due to QW optical phonons. The shift of the Raman intensity maxima towards the situation where the polariton photonic strength is larger is one of the important features that are predicted by this model, and reasonable values for the damping constants are derived. 

The theory in its present form is not able to explain the asymmetry of the observed Raman scans (see Fig.~\ref{fig4}). It might be possible that inelastic mechanisms of decay of the middle polariton in the lower one, not included here, might lead to an improvement of the comparison with experiment. Nevertheless, we believe that the presented Raman scattering theory based on a Green's function approach, allows a consistent introduction of the damping mechanisms, and provides an important step forward in understanding this fundamental and interesting aspect of the inelastic light-matter interaction processes in optical microcavities.

\begin{acknowledgments}
%A.B. thanks jointly CNEA and CONICET from Argentina for financial support.
This work was partially supported by PIP 5254 of CONICET and PICT 2006/483
of ANPCyT (A.A.A.), and ECOS-SUD (A.F. \& B.J.).
\end{acknowledgments}

%\bibliographystyle{apsrev}
%\bibliography{bibliography}

\begin{thebibliography}{37}
\expandafter\ifx\csname natexlab\endcsname\relax\def\natexlab#1{#1}\fi
\expandafter\ifx\csname bibnamefont\endcsname\relax
  \def\bibnamefont#1{#1}\fi
\expandafter\ifx\csname bibfnamefont\endcsname\relax
  \def\bibfnamefont#1{#1}\fi
\expandafter\ifx\csname citenamefont\endcsname\relax
  \def\citenamefont#1{#1}\fi
\expandafter\ifx\csname url\endcsname\relax
  \def\url#1{\texttt{#1}}\fi
\expandafter\ifx\csname urlprefix\endcsname\relax\def\urlprefix{URL }\fi
\providecommand{\bibinfo}[2]{#2}
\providecommand{\eprint}[2][]{\url{#2}}

\bibitem[{\citenamefont{Burstein and Weisbuch}(1995)}]{Book-ConfElAndPhot(95)}
\bibinfo{editor}{\bibfnamefont{E.}~\bibnamefont{Burstein}} \bibnamefont{and}
  \bibinfo{editor}{\bibfnamefont{C.}~\bibnamefont{Weisbuch}}, eds.,
  \emph{\bibinfo{title}{Confined Electrons and Photon: New Physics and
  Applications}} (\bibinfo{publisher}{Plenum Press}, \bibinfo{address}{New
  York}, \bibinfo{year}{1995}).

\bibitem[{\citenamefont{Kavokin and
  Malpuech}(2003)}]{Libro-CavityPolaritons(03)}
\bibinfo{author}{\bibfnamefont{A.}~\bibnamefont{Kavokin}} \bibnamefont{and}
  \bibinfo{author}{\bibfnamefont{G.}~\bibnamefont{Malpuech}},
  \emph{\bibinfo{title}{Cavity Polaritons}} (\bibinfo{publisher}{Elsevier},
  \bibinfo{address}{Amsterdam}, \bibinfo{year}{2003}).

\bibitem[{Sem(2003)}]{SemicondSciTech18-issue10(03)}
\bibinfo{journal}{Semicond. Sci. Technol.} \textbf{\bibinfo{volume}{18}},
  \bibinfo{pages}{S279} (\bibinfo{year}{2003}), \bibinfo{note}{special issue on
  microcavities, J. J. Baumberg and L. Vi\~na, guest editors.}

\bibitem[{JPh(2006)}]{JPhysCondensMatter18-SpecialIssue}
\bibinfo{journal}{J. Phys.: Condens. Matter} \textbf{\bibinfo{volume}{16}},
  \bibinfo{pages}{S3549} (\bibinfo{year}{2006}), \bibinfo{note}{(Special
  issue)}.

\bibitem[{\citenamefont{Kasprzak et~al.}(2006)\citenamefont{Kasprzak, Richard,
  Kundermann, Baas, Jeambrun, Keeling, Marchetti, Szyma\'nska, Andr\'e, Staehli
  et~al.}}]{Kasprzak-Nature443-409(06)}
\bibinfo{author}{\bibfnamefont{J.}~\bibnamefont{Kasprzak}},
  \bibinfo{author}{\bibfnamefont{M.}~\bibnamefont{Richard}},
  \bibinfo{author}{\bibfnamefont{S.}~\bibnamefont{Kundermann}},
  \bibinfo{author}{\bibfnamefont{A.}~\bibnamefont{Baas}},
  \bibinfo{author}{\bibfnamefont{P.}~\bibnamefont{Jeambrun}},
  \bibinfo{author}{\bibfnamefont{J.~M.~J.} \bibnamefont{Keeling}},
  \bibinfo{author}{\bibfnamefont{F.~M.} \bibnamefont{Marchetti}},
  \bibinfo{author}{\bibfnamefont{M.~H.} \bibnamefont{Szyma\'nska}},
  \bibinfo{author}{\bibfnamefont{R.}~\bibnamefont{Andr\'e}},
  \bibinfo{author}{\bibfnamefont{J.~L.} \bibnamefont{Staehli}},
  \bibnamefont{et~al.}, \bibinfo{journal}{Nature}
  \textbf{\bibinfo{volume}{443}}, \bibinfo{pages}{409} (\bibinfo{year}{2006}).

\bibitem[{\citenamefont{Balili et~al.}(2007)\citenamefont{Balili, Hartwell,
  Snoke, Pfeiffer, and West}}]{Balili-Science316-1007(07)}
\bibinfo{author}{\bibfnamefont{R.}~\bibnamefont{Balili}},
  \bibinfo{author}{\bibfnamefont{V.}~\bibnamefont{Hartwell}},
  \bibinfo{author}{\bibfnamefont{D.}~\bibnamefont{Snoke}},
  \bibinfo{author}{\bibfnamefont{L.}~\bibnamefont{Pfeiffer}}, \bibnamefont{and}
  \bibinfo{author}{\bibfnamefont{K.}~\bibnamefont{West}},
  \bibinfo{journal}{Science} \textbf{\bibinfo{volume}{316}},
  \bibinfo{pages}{1007} (\bibinfo{year}{2007}).

\bibitem[{\citenamefont{Deng et~al.}(2006)\citenamefont{Deng, Press,
  G\"otzinger, Solomon, Hey, Ploog, and Yamamoto}}]{Deng-PRL97-146402(06)}
\bibinfo{author}{\bibfnamefont{H.}~\bibnamefont{Deng}},
  \bibinfo{author}{\bibfnamefont{D.}~\bibnamefont{Press}},
  \bibinfo{author}{\bibfnamefont{S.}~\bibnamefont{G\"otzinger}},
  \bibinfo{author}{\bibfnamefont{G.~S.} \bibnamefont{Solomon}},
  \bibinfo{author}{\bibfnamefont{R.}~\bibnamefont{Hey}},
  \bibinfo{author}{\bibfnamefont{L.~H.} \bibnamefont{Ploog}}, \bibnamefont{and}
  \bibinfo{author}{\bibfnamefont{Y.}~\bibnamefont{Yamamoto}},
  \bibinfo{journal}{Phys. Rev. Lett.} \textbf{\bibinfo{volume}{97}},
  \bibinfo{pages}{146402} (\bibinfo{year}{2006}).

\bibitem[{\citenamefont{Imamo\={g}lu and Ram}(1996)}]{Imamoglu-PLA214-193(96)}
\bibinfo{author}{\bibfnamefont{A.}~\bibnamefont{Imamo\={g}lu}}
  \bibnamefont{and} \bibinfo{author}{\bibfnamefont{R.~J.} \bibnamefont{Ram}},
  \bibinfo{journal}{Phys. Lett. A} \textbf{\bibinfo{volume}{214}},
  \bibinfo{pages}{196} (\bibinfo{year}{1996}).

\bibitem[{\citenamefont{Senellart and
  Bloch}(1999)}]{Senellart-Bloch-PRL82-1233(99)}
\bibinfo{author}{\bibfnamefont{P.}~\bibnamefont{Senellart}} \bibnamefont{and}
  \bibinfo{author}{\bibfnamefont{J.}~\bibnamefont{Bloch}},
  \bibinfo{journal}{Phys. Rev. Lett.} \textbf{\bibinfo{volume}{82}},
  \bibinfo{pages}{1233} (\bibinfo{year}{1999}).

\bibitem[{\citenamefont{Dang et~al.}(1998)\citenamefont{Dang, Heger, Andr\'e,
  B{\oe}uf, and Romestain}}]{Dang-PRL81-3920(98)}
\bibinfo{author}{\bibfnamefont{L.~S.} \bibnamefont{Dang}},
  \bibinfo{author}{\bibfnamefont{D.}~\bibnamefont{Heger}},
  \bibinfo{author}{\bibfnamefont{R.}~\bibnamefont{Andr\'e}},
  \bibinfo{author}{\bibfnamefont{F.}~\bibnamefont{Boeuf}}, \bibnamefont{and}
  \bibinfo{author}{\bibfnamefont{R.}~\bibnamefont{Romestain}},
  \bibinfo{journal}{Phys. Rev. Lett.} \textbf{\bibinfo{volume}{81}},
  \bibinfo{pages}{3920} (\bibinfo{year}{1998}).

\bibitem[{\citenamefont{Deveaud}(2006)}]{Book-PhysicsofSemiconductorMicrocavit%
ies(06)}
\bibinfo{editor}{\bibfnamefont{B.}~\bibnamefont{Deveaud}}, ed.,
  \emph{\bibinfo{title}{Physics of Semiconductor Microcavities: From
  Fundamentals to Nanoscale Devices}} (\bibinfo{publisher}{Wiley-VCH Verlag},
  \bibinfo{address}{Berlin}, \bibinfo{year}{2006}).

\bibitem[{\citenamefont{de~Lima~Jr. et~al.}(2006)\citenamefont{de~Lima~Jr.,
  van~der Poel, Santos, and Hvam}}]{deLima-PRL7545326(06)}
\bibinfo{author}{\bibfnamefont{M.~M.} \bibnamefont{de~Lima~Jr.}},
  \bibinfo{author}{\bibfnamefont{M.}~\bibnamefont{van~der Poel}},
  \bibinfo{author}{\bibfnamefont{P.~V.} \bibnamefont{Santos}},
  \bibnamefont{and} \bibinfo{author}{\bibfnamefont{J.~M.} \bibnamefont{Hvam}},
  \bibinfo{journal}{Phys. Rev. Lett.} \textbf{\bibinfo{volume}{97}},
  \bibinfo{pages}{045501} (\bibinfo{year}{2006}).
  

\bibitem[{\citenamefont{Weisbuch et~al.}(1992)\citenamefont{Weisbuch, Nishioka,
  Ishikawa, and Arakawa}}]{Weisbuch-PRL67-3314(92)}
\bibinfo{author}{\bibfnamefont{C.}~\bibnamefont{Weisbuch}},
  \bibinfo{author}{\bibfnamefont{M.}~\bibnamefont{Nishioka}},
  \bibinfo{author}{\bibfnamefont{A.}~\bibnamefont{Ishikawa}}, \bibnamefont{and}
  \bibinfo{author}{\bibfnamefont{Y.}~\bibnamefont{Arakawa}},
  \bibinfo{journal}{Phys. Rev. Lett.} \textbf{\bibinfo{volume}{69}},
  \bibinfo{pages}{3314} (\bibinfo{year}{1992}).

\bibitem[{\citenamefont{Cardona}(1982)}]{Cardona-LightScattering-II}
\bibinfo{author}{\bibfnamefont{M.}~\bibnamefont{Cardona}}, in
  \emph{\bibinfo{booktitle}{Light Scattering in Solids II}}, edited by
  \bibinfo{editor}{\bibfnamefont{M.}~\bibnamefont{Cardona}} \bibnamefont{and}
  \bibinfo{editor}{\bibfnamefont{G.}~\bibnamefont{G\"untherodt}}
  (\bibinfo{publisher}{Springer-Verlag}, \bibinfo{address}{Heidelberg},
  \bibinfo{year}{1982}), vol.~\bibinfo{volume}{50} of
  \emph{\bibinfo{series}{Topics in Applied Physics}}.

\bibitem[{\citenamefont{Fainstein et~al.}(2001)\citenamefont{Fainstein, Trigo,
  Oliva, Jusserand, Freixanet, and Thierry-Mieg}}]{Fainstein-PRL86-3411(01)}
\bibinfo{author}{\bibfnamefont{A.}~\bibnamefont{Fainstein}},
  \bibinfo{author}{\bibfnamefont{M.}~\bibnamefont{Trigo}},
  \bibinfo{author}{\bibfnamefont{D.}~\bibnamefont{Oliva}},
  \bibinfo{author}{\bibfnamefont{B.}~\bibnamefont{Jusserand}},
  \bibinfo{author}{\bibfnamefont{T.}~\bibnamefont{Freixanet}},
  \bibnamefont{and}
  \bibinfo{author}{\bibfnamefont{V.}~\bibnamefont{Thierry-Mieg}},
  \bibinfo{journal}{Phys. Rev. Lett.} \textbf{\bibinfo{volume}{86}},
  \bibinfo{pages}{3411} (\bibinfo{year}{2001}).

\bibitem[{\citenamefont{Fainstein et~al.}(1997)\citenamefont{Fainstein,
  Jusserand, and Thierry-Mieg}}]{Fainstein-PRL78-1576(97)}
\bibinfo{author}{\bibfnamefont{A.}~\bibnamefont{Fainstein}},
  \bibinfo{author}{\bibfnamefont{B.}~\bibnamefont{Jusserand}},
  \bibnamefont{and}
  \bibinfo{author}{\bibfnamefont{V.}~\bibnamefont{Thierry-Mieg}},
  \bibinfo{journal}{Phys. Rev. Lett} \textbf{\bibinfo{volume}{78}},
  \bibinfo{pages}{1576} (\bibinfo{year}{1997}).

\bibitem[{\citenamefont{Tribe et~al.}(1997)\citenamefont{Tribe, Baxter,
  Skolnick, Mowbray, Fisher, and Roberts}}]{Tribe-PRB56-12429(97)}
\bibinfo{author}{\bibfnamefont{W.~R.} \bibnamefont{Tribe}},
  \bibinfo{author}{\bibfnamefont{D.}~\bibnamefont{Baxter}},
  \bibinfo{author}{\bibfnamefont{M.~S.} \bibnamefont{Skolnick}},
  \bibinfo{author}{\bibfnamefont{D.~J.} \bibnamefont{Mowbray}},
  \bibinfo{author}{\bibfnamefont{T.~A.} \bibnamefont{Fisher}},
  \bibnamefont{and} \bibinfo{author}{\bibfnamefont{J.~S.}
  \bibnamefont{Roberts}}, \bibinfo{journal}{Phys. Rev. B}
  \textbf{\bibinfo{volume}{56}}, \bibinfo{pages}{12429} (\bibinfo{year}{1997}).

\bibitem[{\citenamefont{Fainstein et~al.}(1998)\citenamefont{Fainstein,
  Jusserand, and Andr\'e}}]{Fainstein-PRB57-R9439(98)}
\bibinfo{author}{\bibfnamefont{A.}~\bibnamefont{Fainstein}},
  \bibinfo{author}{\bibfnamefont{B.}~\bibnamefont{Jusserand}},
  \bibnamefont{and} \bibinfo{author}{\bibfnamefont{R.}~\bibnamefont{Andr\'e}},
  \bibinfo{journal}{Phys. Rev. B} \textbf{\bibinfo{volume}{57}},
  \bibinfo{pages}{R9439} (\bibinfo{year}{1998}).

\bibitem[{\citenamefont{Bruchhausen et~al.}(2003)\citenamefont{Bruchhausen,
  Fainstein, Jusserand, and Andr\'e}}]{Bruchhausen-PRB68-205326(03)}
\bibinfo{author}{\bibfnamefont{A.}~\bibnamefont{Bruchhausen}},
  \bibinfo{author}{\bibfnamefont{A.}~\bibnamefont{Fainstein}},
  \bibinfo{author}{\bibfnamefont{B.}~\bibnamefont{Jusserand}},
  \bibnamefont{and} \bibinfo{author}{\bibfnamefont{R.}~\bibnamefont{Andr\'e}},
  \bibinfo{journal}{Phys. Rev. B} \textbf{\bibinfo{volume}{68}},
  \bibinfo{pages}{205326} (\bibinfo{year}{2003}).

\bibitem[{\citenamefont{Stevenson et~al.}(2003)\citenamefont{Stevenson,
  Astratov, Skolnick, Roberts, and Hill}}]{Stevenson-PRB67-81301(R)(03)}
\bibinfo{author}{\bibfnamefont{R.~M.} \bibnamefont{Stevenson}},
  \bibinfo{author}{\bibfnamefont{V.~N.} \bibnamefont{Astratov}},
  \bibinfo{author}{\bibfnamefont{M.~S.} \bibnamefont{Skolnick}},
  \bibinfo{author}{\bibfnamefont{J.~S.} \bibnamefont{Roberts}},
  \bibnamefont{and} \bibinfo{author}{\bibfnamefont{G.}~\bibnamefont{Hill}},
  \bibinfo{journal}{Phys. Rev. B} \textbf{\bibinfo{volume}{67}},
  \bibinfo{pages}{081301(R)} (\bibinfo{year}{2003}).

\bibitem[{\citenamefont{Weisbuch and
  Ulbrich}(19??)}]{Weisbuch-LightScattering-III}
\bibinfo{author}{\bibfnamefont{C.}~\bibnamefont{Weisbuch}} \bibnamefont{and}
  \bibinfo{author}{\bibfnamefont{R.~G.} \bibnamefont{Ulbrich}}, in
  \emph{\bibinfo{booktitle}{Light Scattering in Solids III}}, edited by
  \bibinfo{editor}{\bibfnamefont{M.}~\bibnamefont{Cardona}} \bibnamefont{and}
  \bibinfo{editor}{\bibfnamefont{G.}~\bibnamefont{G\"untherodt}}
  (\bibinfo{publisher}{Springer-Verlag}, \bibinfo{address}{Heidelberg},
  \bibinfo{year}{19??}), vol.~\bibinfo{volume}{51} of
  \emph{\bibinfo{series}{Topics in Applied Physics}}.

\bibitem[{\citenamefont{Bruchhausen et~al.}(2005)\citenamefont{Bruchhausen,
  Fainstein, and Jusserand}}]{Bruchhausen-AIPCP772-1117(05)}
\bibinfo{author}{\bibfnamefont{A.}~\bibnamefont{Bruchhausen}},
  \bibinfo{author}{\bibfnamefont{A.}~\bibnamefont{Fainstein}},
  \bibnamefont{and}
  \bibinfo{author}{\bibfnamefont{B.}~\bibnamefont{Jusserand}},
  \bibinfo{journal}{American Institute of Physics}
  \textbf{\bibinfo{volume}{CP772}}, \bibinfo{pages}{1117}
  (\bibinfo{year}{2005}).

\bibitem[{\citenamefont{Hilario et~al.}(2007)\citenamefont{Hilario,
  Bruchhausen, Lobos, and Aligia}}]{LeonHilario-JPCondMatter19-176210(07)}
\bibinfo{author}{\bibfnamefont{L.~M.} \bibnamefont{Hilario}},
  \bibinfo{author}{\bibfnamefont{A.}~\bibnamefont{Bruchhausen}},
  \bibinfo{author}{\bibfnamefont{A.~M.} \bibnamefont{Lobos}}, \bibnamefont{and}
  \bibinfo{author}{\bibfnamefont{A.~A.} \bibnamefont{Aligia}},
  \bibinfo{journal}{J. Phys: Cond. Matter} \textbf{\bibinfo{volume}{19}},
  \bibinfo{pages}{176210} (\bibinfo{year}{2007}).

\bibitem[{\citenamefont{Hilario et~al.}(2008)\citenamefont{Hilario, Aligia,
  Lobos, and Bruchhausen}}]{LeonHilario-InPress-(07)}
\bibinfo{author}{\bibfnamefont{L.~M.} \bibnamefont{Hilario}},
  \bibinfo{author}{\bibfnamefont{A.~A.} \bibnamefont{Aligia}},
  \bibinfo{author}{\bibfnamefont{A.~M.} \bibnamefont{Lobos}}, \bibnamefont{and}
  \bibinfo{author}{\bibfnamefont{A.}~\bibnamefont{Bruchhausen}},
  \bibinfo{journal}{Superlattices and Microstructures}
  \textbf{\bibinfo{volume}{43}}, \bibinfo{pages}{532} (\bibinfo{year}{2008}).

\bibitem[{\citenamefont{Yu and
  Cardona}(1996)}]{Cardona-FundamentalsOfSemicond(96)}
\bibinfo{author}{\bibfnamefont{P.~Y.} \bibnamefont{Yu}} \bibnamefont{and}
  \bibinfo{author}{\bibfnamefont{M.}~\bibnamefont{Cardona}},
  \emph{\bibinfo{title}{Fundamentals of Semiconductors: Physics and Material
  Properies}} (\bibinfo{publisher}{Springer-Verlag}, \bibinfo{address}{Berlin},
  \bibinfo{year}{1996}).

\bibitem[{\citenamefont{Stanley et~al.}(1996)\citenamefont{Stanley, Houdr\'e,
  Weisbuch, Oesterle, and Ilegems}}]{Stanley-PRB53-10995(96)}
\bibinfo{author}{\bibfnamefont{R.~P.} \bibnamefont{Stanley}},
  \bibinfo{author}{\bibfnamefont{R.}~\bibnamefont{Houdr\'e}},
  \bibinfo{author}{\bibfnamefont{C.}~\bibnamefont{Weisbuch}},
  \bibinfo{author}{\bibfnamefont{U.}~\bibnamefont{Oesterle}}, \bibnamefont{and}
  \bibinfo{author}{\bibfnamefont{M.}~\bibnamefont{Ilegems}},
  \bibinfo{journal}{Phys. Rev. B} \textbf{\bibinfo{volume}{53}},
  \bibinfo{pages}{10995} (\bibinfo{year}{1996}).

\bibitem[{\citenamefont{Jorda et~al.}(1993)\citenamefont{Jorda, R\"ossler, and
  Broido}}]{Jorda-PRB48-1669(93)}
\bibinfo{author}{\bibfnamefont{S.}~\bibnamefont{Jorda}},
  \bibinfo{author}{\bibfnamefont{U.}~\bibnamefont{R\"ossler}},
  \bibnamefont{and} \bibinfo{author}{\bibfnamefont{D.}~\bibnamefont{Broido}},
  \bibinfo{journal}{Phys. Rev. B} \textbf{\bibinfo{volume}{48}},
  \bibinfo{pages}{1669} (\bibinfo{year}{1993}).

\bibitem[{\citenamefont{Bruchhausen et~al.}(2006)\citenamefont{Bruchhausen,
  Fainstein, Jusserand, and Andr\'e}}]{Bruchhausen-PRB73-85305(06)}
\bibinfo{author}{\bibfnamefont{A.}~\bibnamefont{Bruchhausen}},
  \bibinfo{author}{\bibfnamefont{A.}~\bibnamefont{Fainstein}},
  \bibinfo{author}{\bibfnamefont{B.}~\bibnamefont{Jusserand}},
  \bibnamefont{and} \bibinfo{author}{\bibfnamefont{R.}~\bibnamefont{Andr\'e}},
  \bibinfo{journal}{Phys. Rev. B} \textbf{\bibinfo{volume}{73}},
  \bibinfo{pages}{85305} (\bibinfo{year}{2006}).

\bibitem[{\citenamefont{Whittaker et~al.}(1996)\citenamefont{Whittaker,
  Kinsler, Skolnick, Armitage, Afshar, Stunge, and
  Roberts}}]{Whittacker-PRL77-4792(96)}
\bibinfo{author}{\bibfnamefont{D.~M.} \bibnamefont{Whittaker}},
  \bibinfo{author}{\bibfnamefont{P.}~\bibnamefont{Kinsler}},
  \bibinfo{author}{\bibfnamefont{T.~A.}~\bibnamefont{Fisher}},
  \bibinfo{author}{\bibfnamefont{M.~S.} \bibnamefont{Skolnick}},
  \bibinfo{author}{\bibfnamefont{A.}~\bibnamefont{Armitage}},
  \bibinfo{author}{\bibfnamefont{A.~M.} \bibnamefont{Afshar}},
  \bibinfo{author}{\bibfnamefont{M.~D.} \bibnamefont{Stunge}},
  \bibnamefont{and} \bibinfo{author}{\bibfnamefont{J.~S.}
  \bibnamefont{Roberts}}, \bibinfo{journal}{Phys. Rev. Lett.}
  \textbf{\bibinfo{volume}{77}}, \bibinfo{pages}{4792} (\bibinfo{year}{1996}).

\bibitem[{\citenamefont{Whittaker}(1998)}]{Whittacker-PRL80-4791(98)}
\bibinfo{author}{\bibfnamefont{D.~M.} \bibnamefont{Whittaker}},
  \bibinfo{journal}{Phys. Rev. Lett.} \textbf{\bibinfo{volume}{80}},
  \bibinfo{pages}{4791} (\bibinfo{year}{1998}), \bibinfo{note}{and references
  therein}.

\bibitem[{\citenamefont{Shinada and
  Sugano}(1966)}]{Shinada-JPhysSocJap21-1936(66)}
\bibinfo{author}{\bibfnamefont{M.}~\bibnamefont{Shinada}} \bibnamefont{and}
  \bibinfo{author}{\bibfnamefont{S.}~\bibnamefont{Sugano}},
  \bibinfo{journal}{J. Phys. Soc. Japan} \textbf{\bibinfo{volume}{21}},
  \bibinfo{pages}{1936} (\bibinfo{year}{1966}).

\bibitem[{\citenamefont{Chao and Chuang}(1991)}]{Chao-PRB43-6530(91)}
\bibinfo{author}{\bibfnamefont{Calvin~Yi-Ping} \bibnamefont{Chao}} \bibnamefont{and}
  \bibinfo{author}{\bibfnamefont{S.~L.} \bibnamefont{Chuang}},
  \bibinfo{journal}{Phys. Rev. B} \textbf{\bibinfo{volume}{43}},
  \bibinfo{pages}{6530} (\bibinfo{year}{1991}).

\bibitem[{\citenamefont{Mahan}(2000)}]{Mahan-ManyParticlePhysics(00)}
\bibinfo{author}{\bibfnamefont{G.~D.} \bibnamefont{Mahan}},
  \emph{\bibinfo{title}{Many Particle Physics}}
  (\bibinfo{publisher}{Kluver/Plenum}, \bibinfo{address}{New York},
  \bibinfo{year}{2000}).

\bibitem[{\citenamefont{Cho}(1979)}]{Cho-Excitons(79)}
\bibinfo{author}{\bibfnamefont{K.}~\bibnamefont{Cho}}, in
  \emph{\bibinfo{booktitle}{Excitons}}, edited by
  \bibinfo{editor}{\bibfnamefont{K.}~\bibnamefont{Cho}}
  (\bibinfo{publisher}{Springer-Verlag}, \bibinfo{address}{Berlin, Heidelberg},
  \bibinfo{year}{1979}), vol.~\bibinfo{volume}{14} of
  \emph{\bibinfo{series}{Topics in Current Physics}}.

\bibitem[{\citenamefont{Jorda}(1994)}]{Jorda-PRB50-2283(94)}
\bibinfo{author}{\bibfnamefont{S.}~\bibnamefont{Jorda}},
  \bibinfo{journal}{Phys. Rev. B} \textbf{\bibinfo{volume}{50}},
  \bibinfo{pages}{2283} (\bibinfo{year}{1994}).

\bibitem[{\citenamefont{Yeh et~al.}(1976)\citenamefont{Yeh, Yariv, and
  Hong}}]{Yeh-JOptSocAm67-473(76)}
\bibinfo{author}{\bibfnamefont{P.}~\bibnamefont{Yeh}},
  \bibinfo{author}{\bibfnamefont{A.}~\bibnamefont{Yariv}}, \bibnamefont{and}
  \bibinfo{author}{\bibfnamefont{C.-S.} \bibnamefont{Hong}},
  \bibinfo{journal}{J. Opt. Soc. Am.} \textbf{\bibinfo{volume}{67}},
  \bibinfo{pages}{423} (\bibinfo{year}{1976}).

\bibitem[{\citenamefont{Yariv}(1971)}]{Yariv-IntroductionToOpticalElectronics(%
71)}
\bibinfo{author}{\bibfnamefont{A.}~\bibnamefont{Yariv}},
  \emph{\bibinfo{title}{Introduction to optical electronics}}
  (\bibinfo{publisher}{Holt, Rinehart, Winston}, \bibinfo{address}{New York},
  \bibinfo{year}{1971}).

\end{thebibliography}

\end{document}